\documentclass{article}

\usepackage[preprint]{neurips_2026}                  

\usepackage[utf8]{inputenc}
\usepackage[T1]{fontenc}
\usepackage{hyperref}
\usepackage{url}
\usepackage{booktabs}     
\usepackage{float}
\usepackage{amsfonts}
\usepackage{amsmath}
\usepackage{nicefrac}
\usepackage{microtype}
\usepackage{xcolor}
\usepackage{graphicx}     
\usepackage{listings}     
\hypersetup{colorlinks=true, linkcolor=blue!50!black, citecolor=blue!50!black, urlcolor=blue!50!black}

\usepackage{subcaption}
\usepackage{hyperref}
\usepackage{xcolor}
\usepackage{fontawesome5}

\title{stable-envs-v1: Fast and Reproducible Environments with Controllable Variations}

\author{%
  \textbf{Sami BuGhanem}$^{1}$\quad
  \textbf{Francesco Capuano}$^{2}$\quad
  \textbf{Ahitagni Das}$^{3}$\quad
  \textbf{Richard Gao}$^{1}$\quad
  \textbf{Randall Balestriero}$^{1}$\\[4pt]
  {\normalfont
    $^{1}$Brown University\quad
    $^{2}$University of Oxford\quad
    $^{3}$Rice University
  }
  \\[8pt]
    {\normalfont\large
      \setlength{\fboxsep}{6pt}%
      \makebox[\textwidth][c]{%
        \hspace*{-3.5em}%
        \colorbox{gray!8}{%
          \href{https://galilai-group.github.io/stable-envs/project/}{%
            \textcolor{gray!65}{\faGlobe\enspace\texttt{Website}}%
          }%
        }%
        \hspace{6pt}%
        \colorbox{gray!8}{%
          \href{https://github.com/galilai-group/stable-envs}{%
            \textcolor{gray!65}{\faGithub\enspace\texttt{Code}}%
          }%
        }%
      }%
    }
}

\begin{document}
\maketitle

\newcommand{\todo}[1]{\textcolor{red}{[TODO: #1]}}   
\newcommand{\lib}{\texttt{stable-envs}}              
\newcommand{\mytitle}{stable-environments-v1: Fast Environments for Faster World Model Training}

\vspace{-2em}
\begin{abstract}
Training World Models (WMs) relies on processing large amounts of high-dimensional data, which typically needs to be stored, read into memory, and moved from memory to compute units.
In principle, GPU-native rendering avoids this movement of large image tensors from memory to compute cores, allowing for better GPU utilization, and training speed-ups.
Rendering on-demand also allows to control perceptual appearance through the application of visual variations on-the-fly, allowing for higher diversity in training.
To this end, we introduce \lib, an open-source library for fast, GPU-native rendering allowing to apply controllable variations on multiple simulation environments.
By rendering image observations directly on GPU, \lib~allows end-to-end training speed-ups of up to $1.33\times$, with no loss of performance.
Besides faster training runs, \lib~also enables reproducible, large-scale evaluations through batched physics while maintaining full determinism.
\lib~currently supports 80 simulation environments across multiple CPU- and GPU-backends, and is openly available at: \url{https://github.com/galilai-group/stable-envs}.
\end{abstract}

\begin{figure}[h!]
  \centering
  \includegraphics[width=0.9\linewidth]{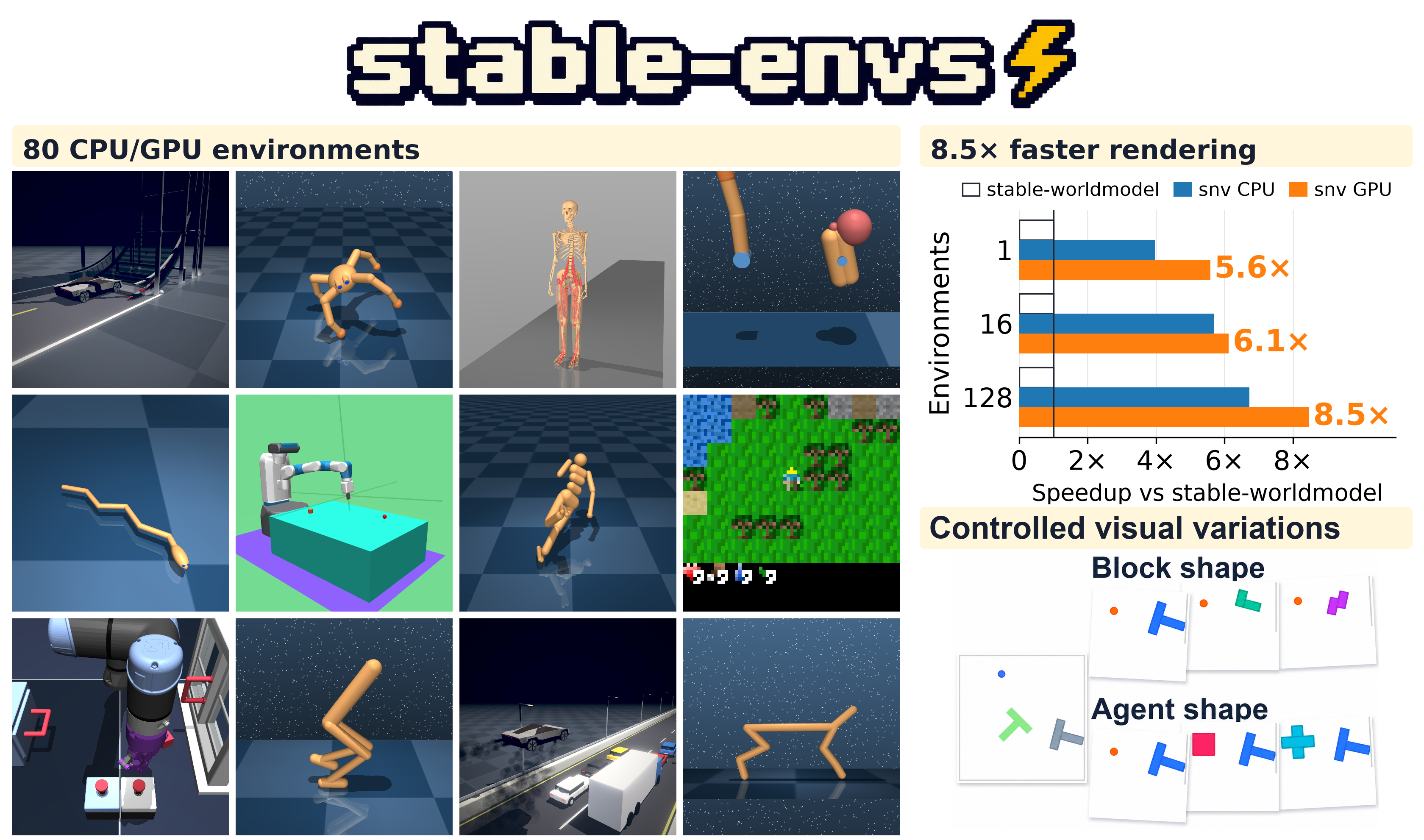}
  \caption{\lib~is an open-source library for batched, on-GPU rendering of simulation environments. It supports a growing number of environments, enables up to 8.5x speedups in rendering, and allows controlled visual variations given the same underlying physical state.}
  \label{fig:fig1}
\end{figure}

\section{Introduction}

Pixel-based World Models (WMs) learn environmental dynamics solely from visual observations.
In practice, training WMs from image frames requires efficiently processing large amounts of high-dimensional image data, collected (1) online, interacting with an environment~\citep{ha2018world, hafner2019learning, hafner2025mastering}, (2) offline, via a behavior policy used to collect trajectories~\citep{zhou2024dino, sobal2026learning, maes2026leworldmodel}, or (3) both.
Regardless of its source, image data can be challenging to process: even compact WMs trained on medium-sized, task-specific datasets such as~\citet[LeWM]{maes2026leworldmodel}, can require processing \(\sim\)200GB of image data.\footnote{1 byte per pixel \texttt{@uint8}, 224x224 frames, 20k trajectories of 200 steps each on average.}.
Storing and processing such volumes for simple environments (e.g., maze navigation) can be challenging in practice, and potentially prohibitive for complex environments, e.g. high-fidelity robotics simulators, e.g.~\citet{mittal2025isaac}.

To mitigate these storage requirements, open-source libraries built for training models from pixel-space trajectories often store visual observations as compressed videos, which are then decoded into individual frames before or during training, e.g.,~\citet{cadene2026lerobot}.
While effective for storage, video decoding introduces additional overhead: loading, decoding, and storing individual frames ahead of training can take several hours for large datasets.
Streaming decoded frames concurrently with model updates can avoid this time-consuming pre-processing step, but requires carefully balancing frame production and consumption to prevent GPU under-utilization.
Crucially, both approaches typically still require transferring decoded frames from CPU to GPU, introducing additional data-movement overhead.

For WMs trained \emph{in simulation}, rendering observations (1) on-the-fly and (2) directly on the GPU avoids storing large image datasets and transferring image tensors between CPU and GPU.
However, rendering can itself become a training bottleneck, making frame throughput critical for efficient training.
Existing libraries typically achieve high throughput through CPU-based environment vectorization and parallelization~\citep{raffin2021stable,towers2026gymnasium,weng2022envpool}.
While effective at scaling environment execution, these approaches do not necessarily accelerate individual environments and can incur substantial overhead when generating and transferring large image tensors to the GPU.
Other frameworks, such as Isaac Lab~\citep{mittal2025isaac}, ManiSkill3~\citep{tao2025demonstrating}, and MuJoCo Playground~\citep{zakka2025mujoco}, leverage GPU-accelerated simulation and rendering to achieve high throughput across many environments on the GPU.
However, these frameworks often rely on specialized simulation backends and hardware requirements, limiting portability across research workflows---e.g., Isaac Lab's GPU renderer is not supported by Hopper-class NVIDIA GPUs (e.g., H100) nor macOS.

To mitigate these scalability and accessibility issues, we introduce \lib, a lightweight, research-friendly library, designed for fast and reproducible GPU-native rendering across multiple simulation environments.
For each supported environment, we port or re-implement the original physics simulation and expose its complete internal state, enabling (1) fast rendering (2) exact state saving and restoration for reproducibility and (3) controllable visual variations behind a single, shared API.
In summary, our contributions are:
\begin{itemize}
    \item \textbf{Fast, reproducible, and controllable environment rendering.} Through optimizations such as batched rendering and custom GPU kernels, \lib~achieves up to 8.5x higher environment throughput than CPU-native baselines. By exposing the full simulator state, \lib~also enables exact state restoration and independent sampling of visual factors of variation, allowing fully-reproducible experiments under controlled visual changes.

    \item \textbf{A comprehensive, GPU-native simulation library.} \lib~currently supporting 80 simulation environments used for world-model research. All environments support batched physics stepping and rendering on the GPU, allowing for efficient training and evaluation of pixel-based WMs.

    \item \textbf{Accelerated end-to-end world-model training.} Thanks to on-the-fly and on-GPU rendering, \lib~enables up to 1.33x faster training of WMs through \texttt{stable-worldmodel}~\citep{maes2026stable}, with no degradation in downstream planning performance.
\end{itemize}
\noindent
\begin{minipage}[t]{0.48\linewidth}
\begin{lstlisting}[
  language=Python,
  basicstyle=\ttfamily\scriptsize,
  breaklines=true,
  caption={Step multiple batched environments.},
  label={lst:simulate}
]
import stable_envs as snv

env = snv.PushT(num_envs=256, device="cuda")
env.reset(seed=0)

# random [B, T, A] actions 
actions = env.action_space.sample()

# Batched-physics, optional rendering
with_frames = env.step(actions, frames=slice(None))
no_frames = env.step(actions, frames=None)  # 1.7x
\end{lstlisting}
\end{minipage}\hfill%
\begin{minipage}[t]{0.48\linewidth}
\begin{lstlisting}[
  language=Python,
  basicstyle=\ttfamily\scriptsize,
  breaklines=true,
  caption={Apply controlled visual variations.},
  label={lst:render}
]
import stable_envs as snv

env = snv.PushT(num_envs=256, device="cuda")
env.reset(seed=0)

before = env.render()[0]
# Change the the T-block's RGB color
env.set_variation(
    {"block.color": [230, 80, 60]}
)
after = env.render()[0]  # after != before
\end{lstlisting}
\end{minipage}

\section{\lib{}: An Overview}
\label{sec:overview}

We interface a total of 80 traditional (e.g., \texttt{TwoRoom}, \texttt{PushT}, etc.) and novel (e.g., \texttt{RoadSter}) simulation environments from multiple environment classes (Figure~\ref{fig:all-envs}, Table~\ref{tab:coverage}).
Each environment class owns its low-level logic, while \lib~implements higher-level factor-of-variations and batched calls to stepping methods, each validated against source.
Importantly, all environment instances step together through one shared simulation adapter. 
On CPU device, this single adapter runs vectorized code or distributes environments across worker processes, whereas on GPU it runs batched MuJoCo Warp, MJX or Pytorch kernels, depending on the environment.
Thus, each environment in \lib~can be used to simulate a batch of \( B \) independent environments, each providing two main methods over an explicit environment state: dynamics stepping (Code ex.~\ref{lst:simulate}), and rendering (Code ex.~\ref{lst:render}).
All environments currently supported in \lib~are listed in Table~\ref{tab:coverage}.
In practice, environments are first \texttt{reset}, returning a \( B -\)batched state, and then \texttt{step} through a (batched) action tensor, using either single actions ($[B,A]$) or action chunks ($[B,T,A]$). 
Crucially, \texttt{step} exposes the intermediate states and the optionally-rendered frames, decoupling environment stepping from rendering.
Similarly, the \texttt{render} routine can be used to render the environment's internal state into \( B \) RGB frames at an arbitrary resolution.
Importantly, state payloads store the low-level deterministic and stochastic information about the environment, so that dynamics are fully reproducible given (1) the environment class and (2) prerecorded states.
Effectively, \lib~allows to store datasets using only implicit states that are then rendered on-the-fly, instead of storing explicit high-dimensional image frames, by exploiting the bijection between the environment's state and its visual outlook.

\paragraph{Performance gains}

\begin{figure}[t]
  \vspace{-1em}
  \centering
  \begin{subfigure}[t]{0.49\linewidth}
    \centering
    \includegraphics[width=\linewidth]{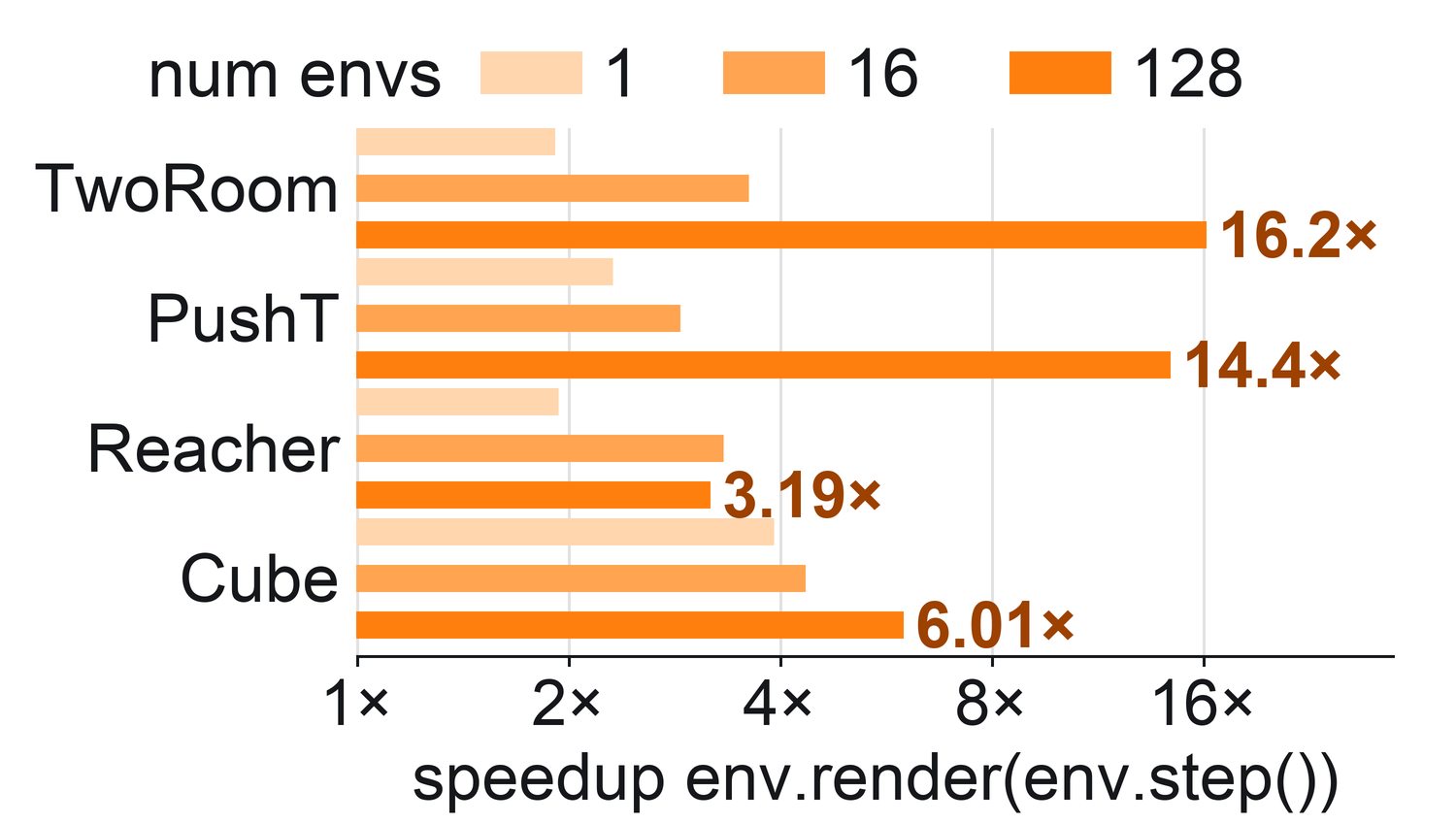}
    \caption{\lib~speeds up combined physics and rendering (\texttt{env.render(env.step())}) compared to \texttt{stable-worldmodel}.}
    \label{fig:env-speedup}
  \end{subfigure}\hfill
  \begin{subfigure}[t]{0.49\linewidth}
    \centering
    \includegraphics[width=\linewidth]{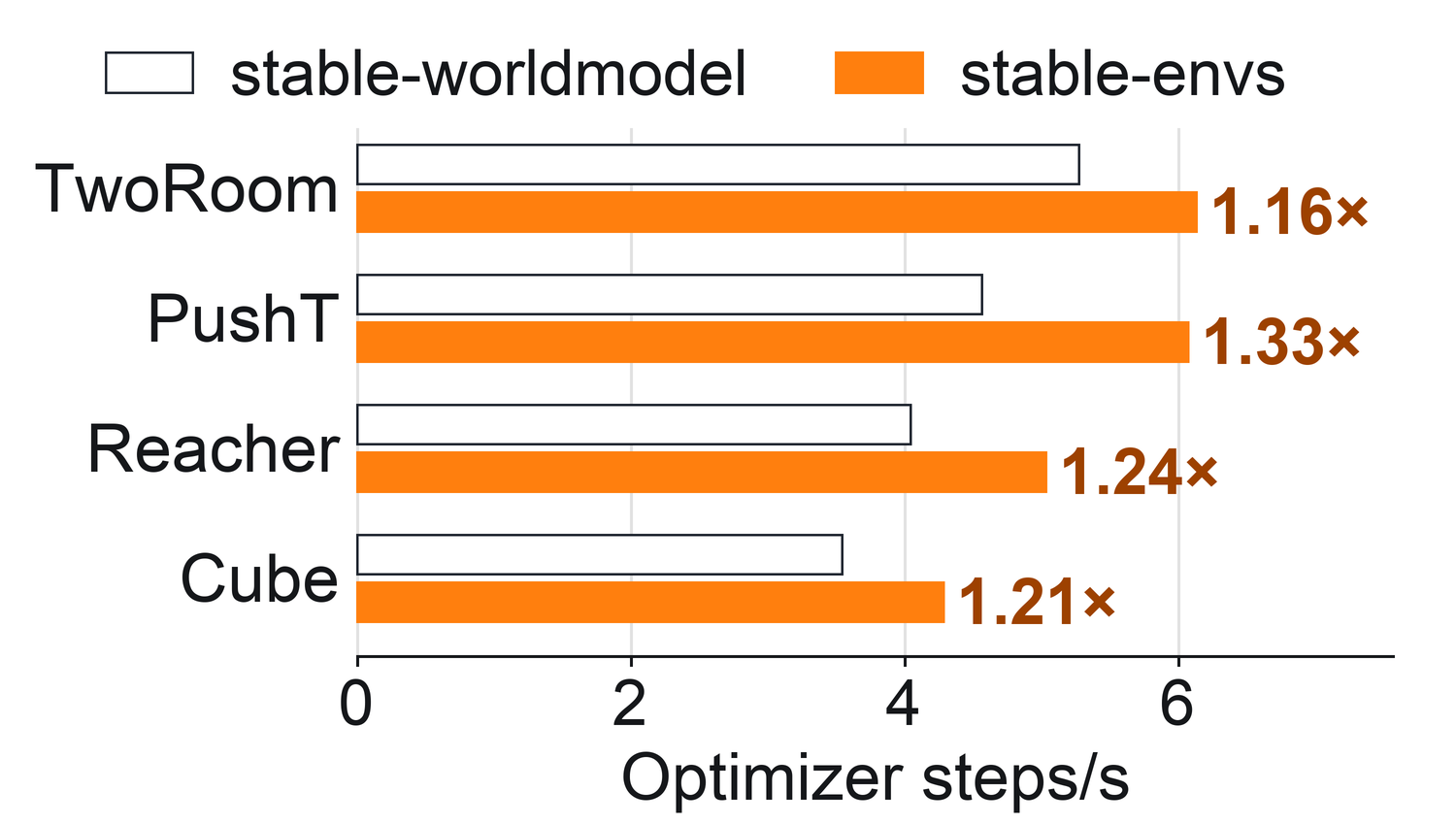}
    \caption{Thanks to on-GPU rendering, \lib~enables significant end-to-end speedups of training across multiple environments.}
    \label{fig:e2e-training-speedup}
  \end{subfigure}
  \caption{Simulator and training performance with \lib.}
  \label{fig:simulator-and-training-speedups}
\end{figure}

\lib~takes advantage of several techniques for rendering speedups.
First, \lib~can render static scenes once per each configuration across the batch, and only draw moving bodies on top for each frame.
Second, \lib~relies on custom kernels to draw frames per each environment, using vectorized kernels or process pools for CPU rendering, and MuJoCo Warp and \texttt{nvdiffrast} for native GPU rendering across compute platforms.
On-GPU rendering is a particularly compelling approach as it allows to directly write tensors into the same device running training.
Further, \lib's is also inherently faster than its reference implementation (Figure~\ref{fig:env-speedup}), with environment's pure rendering from prerecorded state trajectories being 1.4x--24.1x faster on GPU when rendering multiple states in parallel, e.g., to form a training batch (Appendix~\ref{app:speedups}).
We also assess \lib's speedups separately for (1) pure physics stepping (\texttt{env.step()}) (2) pure frame rendering (\texttt{env.render())}, and (3) both (\texttt{env.render(env.step())}) on both CPU and GPU, and report our results in Table~\ref{tab:speedups-full}.
We directly rely on these speed ups to accelerate the training run of LeWM~\citep{maes2026leworldmodel}, a compact world model trained solely on image observations.
Across four environments (\texttt{TwoRoom}, \texttt{PushT}, \texttt{Reacher} and \texttt{Cube}), training runs are (1) 1.16$\times$--1.33$\times$ faster (Figure~\ref{fig:e2e-training-speedup}) and (2) exhibit higher GPU utilization (Table~\ref{tab:lewm-gpu-util}) than the reference implementation, notably with no changes to the training loop besides on-GPU rendering.
Importantly, these optimized training runs lead to comparable downstream planning performance (Table~\ref{tab:training_speedup_correctness}).
  
\begin{figure}
  \vspace{-1em}
  \centering
  \includegraphics[width=\linewidth]{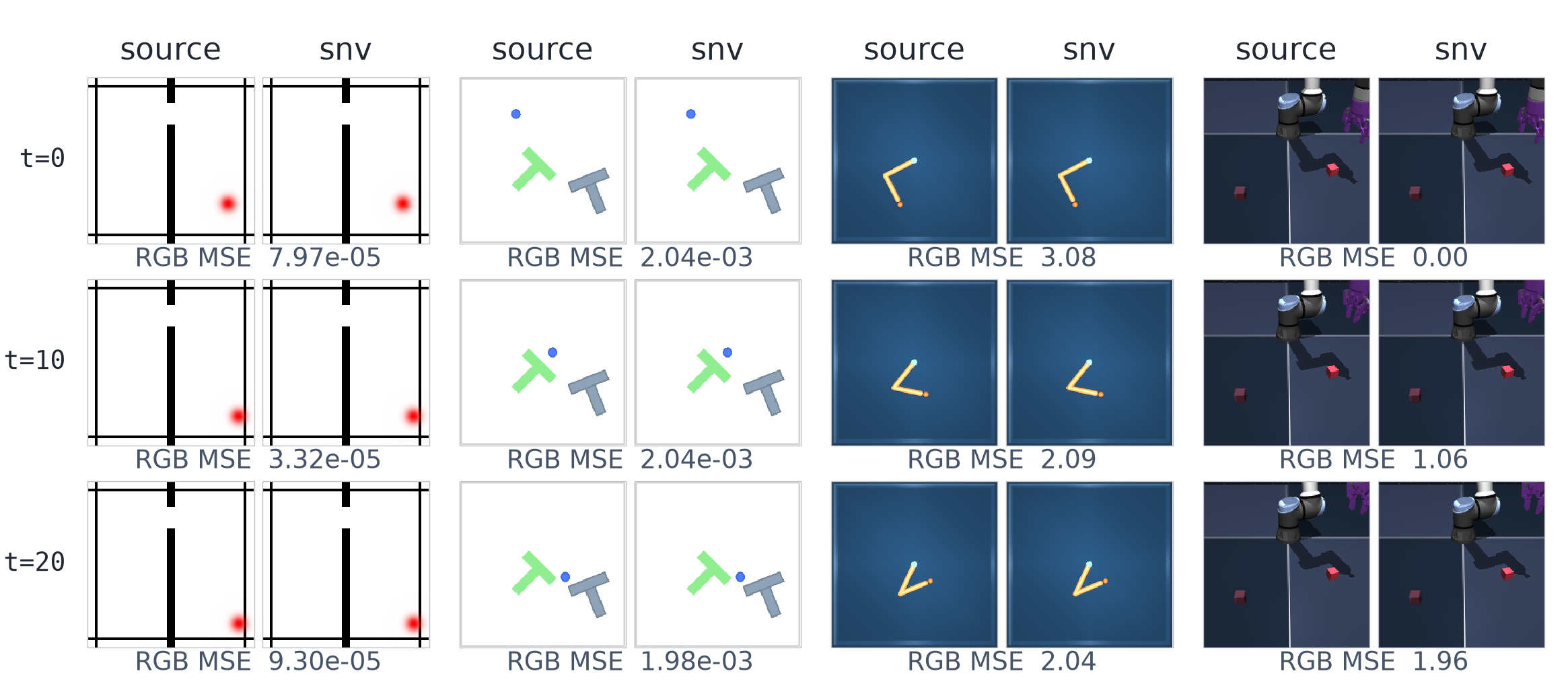}
  \caption{\lib~shows both (1) dynamics and (2) perceptual parity with source implementation despite being optimized for speed. Environments from left to right are: \texttt{TwoRoom}, \texttt{PushT}, \texttt{Reacher} and \texttt{Cube}.}
  \label{fig:parity}
\end{figure}

\paragraph{Environments correctness}
To assess environment correctness, \lib~tests next-state dynamics and rendering parity with source environments separately.
To guarantee parity in the simulated dynamics, we run both the source implementation and \lib's from the same state \( s_0 \) given a shared action sequence \(a_0, a_1, \dots a_{20} \), comparing the resulting trajectories, and report the maximum error in Table~\ref{tab:dynamics-vs-source-error}.
Further, we assess rendering's semantic correctness by testing source and \lib's~rendering of the same underlying state \( s_t \), and report the pixel-space MSE against the source implementation in Table~\ref{tab:rendering-vs-source-error}.
On CPU, all four environments are near pixel-identical, and on GPU the MSE is lower still: \texttt{TwoRoom}, \texttt{PushT} and \texttt{Cube} are almost exact, and only \texttt{Reacher} exhibits a negligible difference of 2.2 (Table~\ref{tab:rendering-vs-source-error}).
Despite the significant speed ups due to \lib's optimized implementation, we do not find significant discrepancies in either dynamics stepping nor visual appearance for rendered scenes, further underscoring \lib's parity with respect to source (Figure~\ref{fig:parity}).
 
 \section{Conclusion}

We introduced \lib, an open-source library for efficient batch rendering of simulated environments designed to speed up research in world models and pixel-based agents.
Currently, \lib~supports over 80 environments, offering end-to-end training speedups of up to 1.33x, and standalone environment speedups of up to 24x for batches of simulation environments.
\lib~achieves this by avoiding moving large image tensors across the CPU/GPU boundary, together with batched rendering, custom rendering kernels, and re-use of pre-rendered static scenery.
\lib~also provides access to controlled visual variations from the same underlying physical state, allowing changes to the visual appearance of all supported environments at runtime.
In addition, \lib~allows datasets to be stored as low-dimensional simulator states and rendered on demand, at different resolutions and under controlled visual variations, during training.
This allows to achieve better GPU utilization, and up to a 1.33x end-to-end training speedup.
We hope \lib{} will help reduce the computational cost of training and evaluating pixel-based WMs and agents, making larger-scale and more reproducible studies in simulated environments accessible to a broader community.

\bibliographystyle{plainnat}
\bibliography{references}

@article{maes2026stable,
  title={stable-worldmodel-v1: Reproducible world modeling research and evaluation},
  author={Maes, Lucas and Lidec, Quentin Le and Haramati, Dan and Massaudi, Nassim and Scieur, Damien and LeCun, Yann and Balestriero, Randall},
  journal={arXiv preprint arXiv:2602.08968},
  year={2026}
}

@article{maes2026leworldmodel,
  title={Leworldmodel: Stable end-to-end joint-embedding predictive architecture from pixels},
  author={Maes, Lucas and Lidec, Quentin Le and Scieur, Damien and LeCun, Yann and Balestriero, Randall},
  journal={arXiv preprint arXiv:2603.19312},
  year={2026}
}

@article{zhou2024dino,
  title={Dino-wm: World models on pre-trained visual features enable zero-shot planning},
  author={Zhou, Gaoyue and Pan, Hengkai and LeCun, Yann and Pinto, Lerrel},
  journal={arXiv preprint arXiv:2411.04983},
  year={2024}
}

@article{sobal2026learning,
  title={Learning from reward-free offline data: A case for planning with latent dynamics models},
  author={Sobal, Uladzislau and Zhang, Wancong and Cho, Kyunghyun and Balestriero, Randall and Rudner, Tim GJ and LeCun, Yann},
  journal={Advances in Neural Information Processing Systems},
  volume={38},
  pages={43905--43941},
  year={2026}
}

@article{hafner2025mastering,
  title={Mastering diverse control tasks through world models},
  author={Hafner, Danijar and Pasukonis, Jurgis and Ba, Jimmy and Lillicrap, Timothy},
  journal={Nature},
  volume={640},
  number={8059},
  pages={647--653},
  year={2025},
  publisher={Nature Publishing Group UK, London}
}

@inproceedings{hafner2019learning,
  title={Learning latent dynamics for planning from pixels},
  author={Hafner, Danijar and Lillicrap, Timothy and Fischer, Ian and Villegas, Ruben and Ha, David and Lee, Honglak and Davidson, James},
  booktitle={International conference on machine learning},
  pages={2555--2565},
  year={2019},
  organization={PMLR}
}

@article{ha2018world,
  title={World models},
  author={Ha, David and Schmidhuber, J{\"u}rgen},
  journal={arXiv preprint arXiv:1803.10122},
  volume={2},
  number={3},
  pages={440},
  year={2018}
}

@article{towers2026gymnasium,
  title={Gymnasium: A standard interface for reinforcement learning environments},
  author={Towers, Mark and Kwiatkowski, Ariel and Balis, John and De Cola, Gianluca and Deleu, Tristan and Goul{\~a}o, Manuel and Andreas, Kallinteris and Krimmel, Markus and Kg, Arjun and Perez-Vicente, Rodrigo and others},
  journal={Advances in Neural Information Processing Systems},
  volume={38},
  year={2026}
}

@article{weng2022envpool,
  title={Envpool: A highly parallel reinforcement learning environment execution engine},
  author={Weng, Jiayi and Lin, Min and Huang, Shengyi and Liu, Bo and Makoviichuk, Denys and Makoviychuk, Viktor and Liu, Zichen and Song, Yufan and Luo, Ting and Jiang, Yukun and others},
  journal={Advances in Neural Information Processing Systems},
  volume={35},
  pages={22409--22421},
  year={2022}
}

@article{mittal2025isaac,
  title={Isaac lab: A gpu-accelerated simulation framework for multi-modal robot learning},
  author={Mittal, Mayank and Roth, Pascal and Tigue, James and Richard, Antoine and Zhang, Octi and Du, Peter and Serrano-Munoz, Antonio and Yao, Xinjie and Zurbr{\"u}gg, Ren{\'e} and Rudin, Nikita and others},
  journal={arXiv preprint arXiv:2511.04831},
  year={2025}
}

@article{tao2025demonstrating,
  title={Demonstrating gpu parallelized robot simulation and rendering for generalizable embodied ai with ManiSkill3},
  author={Tao, S and Xiang, F and Shukla, A and Qin, Y and Hinrichsen, X and Yuan, X and Bao, C and Lin, X and Liu, Y and Chan, TK and others},
  journal={Proceedings of Robotics: Science and Systems,(LosAngeles, CA, USA)},
  year={2025}
}

@article{zakka2025mujoco,
  title={Mujoco playground},
  author={Zakka, Kevin and Tabanpour, Baruch and Liao, Qiayuan and Haiderbhai, Mustafa and Holt, Samuel and Luo, Jing Yuan and Allshire, Arthur and Frey, Erik and Sreenath, Koushil and Kahrs, Lueder A and others},
  journal={arXiv preprint arXiv:2502.08844},
  year={2025}
}

@inproceedings{cadene2026lerobot,
  title={Lerobot: An open-source library for end-to-end robot learning},
  author={Cadene, Remi and Alibert, Simon and Capuano, Francesco and Aractingi, Michel and Zouitine, Adil and Kooijmans, Pepijn and Choghari, Jade and Russi, Martino and Pascal, Caroline and Palma, Steven and others},
  booktitle={International Conference on Learning Representations},
  volume={2026},
  pages={122398--122417},
  year={2026}
}

@article{raffin2021stable,
  title={Stable-baselines3: Reliable reinforcement learning implementations},
  author={Raffin, Antonin and Hill, Ashley and Gleave, Adam and Kanervisto, Anssi and Ernestus, Maximilian and Dormann, Noah},
  journal={Journal of machine learning research},
  volume={22},
  number={268},
  pages={1--8},
  year={2021}
}

\newpage
\appendix

\section{Environments available}

\begin{figure}[H]
    \centering
    \includegraphics[width=\linewidth]{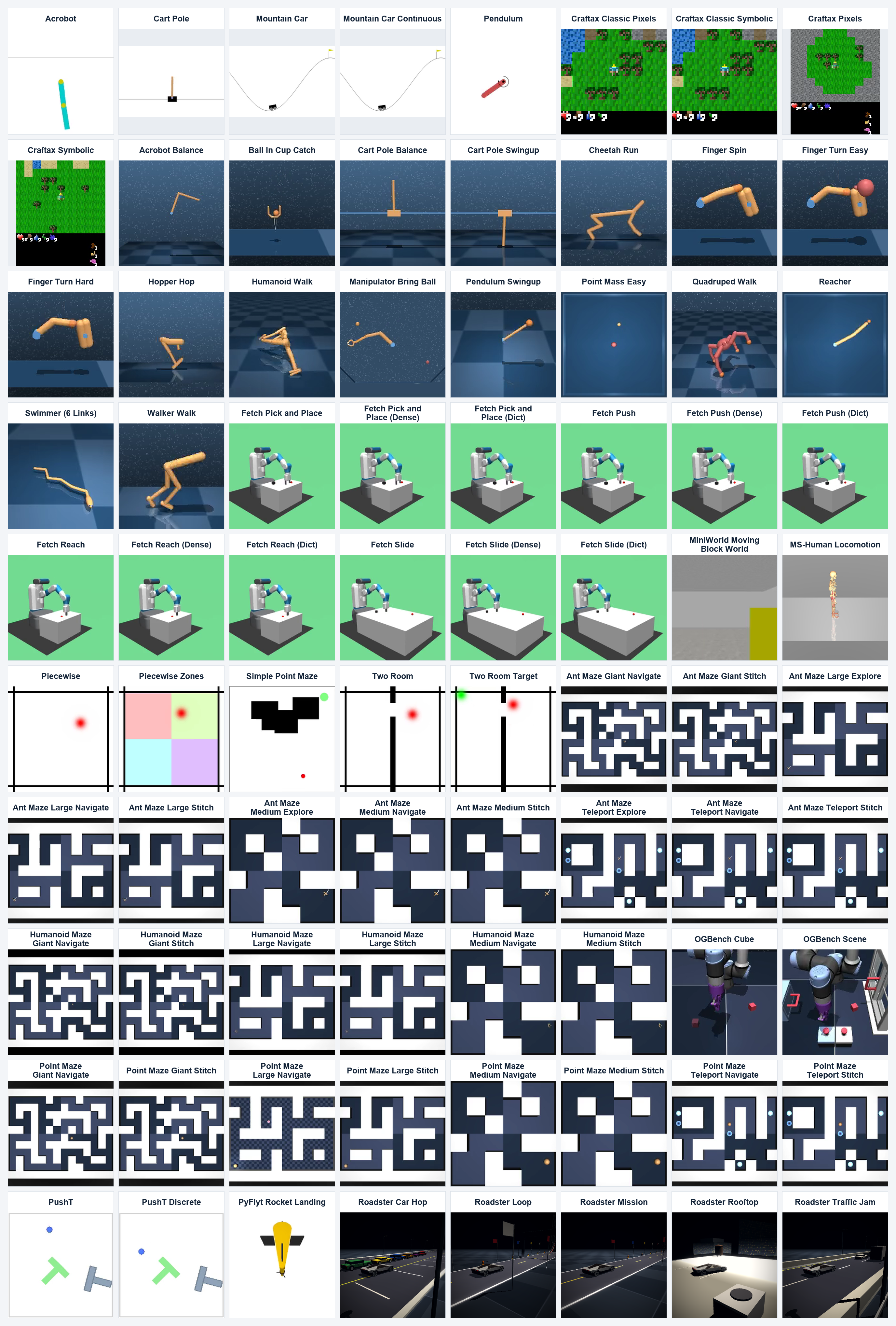}
    \caption{All 80 environments currently supported by \lib.}
    \label{fig:all-envs}
\end{figure}

\begin{table}[H]
\centering
\small
\caption{Environment families in \lib{}.}
\label{tab:coverage}
\begin{tabular}{llr}
  \toprule
  Family & Source & Classes \\
  \midrule
  Classic control & Gymnasium & 5 \\
  DM Control & dm\_control (MuJoCo) & 17 \\
  Fetch & Gymnasium Robotics (MuJoCo) & 12 \\
  OGBench & OGBench (MuJoCo) & 27 \\
  Navigation & stable-worldmodel & 5 \\
  PushT & gym-pusht (Pymunk) & 2 \\
  Roadster & new in \lib{} (MuJoCo) & 5 \\
  MiniWorld & MiniWorld & 1 \\
  PyFlyt & PyFlyt (PyBullet) & 1 \\
  MS-Human & MS-Human-700 (MuJoCo) & 1 \\
  Craftax & Craftax (JAX) & 4 \\
  \bottomrule
\end{tabular}
\end{table}

\section{Benchmark protocol}
\label{app:protocol}
\begin{table}[H]
\centering\small
\caption{Hardware and protocol for the speed and correctness measurements.}
\label{tab:protocol}
\begin{tabular}{ll}
\toprule
GPU & NVIDIA L40S (driver 595.91) \\
CPU & AMD EPYC 9354, 8 pinned cores \\
Frame size & $224\times224$ on both sides \\
Arguments & library defaults \\
Timing & warm-up, then 3 windows of at least 3\,s (5 at $B{=}1$); median \\
\bottomrule
\end{tabular}
\end{table}

\section{Environments speedups}
\label{app:speedups}

\begin{table}[H]
\centering\small
\caption{Speedup of \lib{} over \texttt{stable-worldmodel} (ratio of steps/s or frames/s) at batch sizes $B$ = 1, 16, 128. "Both" denotes \texttt{env.render(env.step())}. GPU physics is compared with \texttt{stable-worldmodel}'s CPU stepping, which often has no GPU path. For GPU rendering, we benchmark against the faster render between \texttt{stable-worldmodel}'s CPU and EGL rendering. \underline{\textbf{Maximum speedups}} per environment and operation, across all devices and batch sizes are highlighted.}
\label{tab:speedups-full}
\resizebox{\linewidth}{!}{%
\begin{tabular}{llccccccccc}
\toprule
& & \multicolumn{3}{c}{\texttt{env.step()}} & \multicolumn{3}{c}{\texttt{env.render()}} & \multicolumn{3}{c}{Both} \\
\cmidrule(lr){3-5}\cmidrule(lr){6-8}\cmidrule(lr){9-11}
Env & Device & 1 & 16 & 128 & 1 & 16 & 128 & 1 & 16 & 128 \\
\midrule

 TwoRoom & CPU & 3.89 & \underline{\textbf{7.90}} & 6.55 & 11.5 & 17.4 & 21.3 & 3.49 & 5.69 & 12.1 \\
          & GPU & 0.307 & 0.841 & 1.77 & 14.2 & 16.0 & \underline{\textbf{24.1}} & 0.887 & 2.23 & \underline{\textbf{13.2}} \\
  PushT   & CPU & 1.36 & 0.96 & \underline{\textbf{2.01}} & 2.94 & 4.02 & 4.13 & 3.74 & 3.44 & 11.7 \\
          & GPU & 0.342 & 0.522 & 1.58 & 3.92 & 4.23 & \underline{\textbf{5.29}} & 1.35 & 2.59 & \underline{\textbf{14.4}} \\
  Reacher & CPU & 2.27 & \underline{\textbf{3.24}} & 3.13 & 18.0 & 18.1 & \underline{\textbf{19.0}} & 8.24 & 8.31 & \underline{\textbf{13.7}} \\
          & GPU & 0.275 & 0.762 & 1.33 & 2.85 & 2.87 & 3.08 & 1.38 & 3.03 & 3.19 \\
  Cube    & CPU & 5.79 & 5.71 & \underline{\textbf{7.61}} & 12.6 & \underline{\textbf{12.8}} & \underline{\textbf{12.8}} & \underline{\textbf{18.0}} & 5.19 & 1.41
  \\
          & GPU & 0.783 & 1.41 & 7.55 & 1.37 & 1.38 & 1.44 & 0.951 & 2.18 & 6.45 \\
\bottomrule
\end{tabular}}
\end{table}

\section{Training speedups}

\begin{table}[H]
\centering\small
\caption{LeWM training (18M parameters, \texttt{bf16}, batch size 128) throughput (\( \uparrow \)) on one NVIDIA L40S, measured in optimizer steps/s after warmup (steps 201--1000). Inputs are frames rendered online by either \texttt{stable-worldmodel} or \lib{}. Model only denotes training on frames already on the GPU.}
\label{tab:lewm-throughput}
\begin{tabular}{lccc}
\toprule
Env & stable-worldmodel & \lib{} & Model only \\
\midrule
TwoRoom & 5.29 & 6.15 & 6.78 \\
PushT   & 4.58 & 6.09 & 6.77 \\
Reacher & 4.06 & 5.05 & 6.78 \\
Cube    & 3.56 & 4.29 & 6.78 \\
\bottomrule
\end{tabular}
\end{table}

\begin{table}[H]
\centering\small
\caption{GPU utilization (\%) (\( \uparrow \)) during LeWM training, measured as the share of wall time the GPU spends on the training step. Rendering time is excluded.}
\label{tab:lewm-gpu-util}
\begin{tabular}{lcc}
\toprule
Env & stable-worldmodel & \lib{} \\
\midrule
TwoRoom & 78.2 & \textbf{86.3} \\
PushT   & 71.8 & \textbf{86.2} \\
Reacher & 57.9 & \textbf{72.3} \\
Cube    & 52.3 & \textbf{62.2} \\
\bottomrule
\end{tabular}
\end{table}

\begin{table}[H]
\centering\small
\caption{Time-to-1k-training-steps (s) (\( \downarrow \)), measured from the start of the first optimizer step to the end of step 1000 for .}
\label{tab:lewm-training-time}
\begin{tabular}{lcc}
\toprule
Env & stable-worldmodel & \lib{} \\
\midrule
TwoRoom & 234 & \textbf{211} \\
PushT   & 252 & \textbf{221} \\
Reacher & 301 & \textbf{250} \\
Cube    & 330 & \textbf{291} \\
\bottomrule
\end{tabular}
\end{table}

\section{Environment correctness}

\begin{table}[h]
\centering\small
\caption{Dynamics parity: maximum absolute error (\( \downarrow \)) over states executing 20 shared actions, from the same seeded reset.}
\label{tab:dynamics-vs-source-error}
\begin{tabular}{lc}
\toprule
Env & Max error \\
\midrule
TwoRoom & 0 \\
PushT   & $4\times10^{-13}$ \\
Reacher & 0 \\
Cube    & 0 \\
\bottomrule
\end{tabular}
\end{table}

\begin{table}[h]
\centering\small
\caption{Rendering MSE (\( \downarrow \)) against \texttt{stable-worldmodel} (0--255 scale, $224\times224$), mean and max over rendered states.}
\label{tab:rendering-vs-source-error}
\begin{tabular}{lcccc}
\toprule
& \multicolumn{2}{c}{CPU} & \multicolumn{2}{c}{GPU} \\
\cmidrule(lr){2-3}\cmidrule(lr){4-5}
Env & Mean & Max & Mean & Max \\
\midrule
TwoRoom & $2\times10^{-5}$ & $4\times10^{-4}$ & $2\times10^{-5}$ & $4\times10^{-4}$ \\
PushT   & 0 & 0 & 0 & 0 \\
Reacher & 2.20 & 2.86 & 2.16 & 2.75 \\
Cube    & 2.02 & 4.31 & $4\times10^{-7}$ & $3\times10^{-5}$ \\
\bottomrule
\end{tabular}
\end{table}

\section{Training correctness}

\begin{table}[H]
  \centering
  \caption{Downstream planning success rates (\%) (\( \uparrow \)) for world models trained and evaluated using the original environments (Source) or \lib. Values report the average over three evaluation seeds.}
  \label{tab:training_speedup_correctness}
  \begin{tabular}{lcc}
      \toprule
      \textbf{Environment} & Source & \textbf{\lib{}} \\
      \midrule
      TwoRoom & $88.7 $ & $88.0 $ \\
      PushT   & $94.0 $ & $93.3 $ \\
      Reacher & $80.0 $ & $79.3 $ \\
      Cube    & $74.0$ & $71.3 $ \\
      \bottomrule
  \end{tabular}
\end{table}

\end{document}